\documentclass[english]{article}
\usepackage[T1]{fontenc}
\usepackage[latin9]{inputenc}
\usepackage{geometry}
\usepackage{color}
\usepackage{array}
\date{}
\makeatletter

\providecommand{\tabularnewline}{\\}

\usepackage{cite,times}

\makeatother

\usepackage{babel}
\begin{document}

\title{Quantum e-commerce: A comparative study of possible protocols for
online shopping and other tasks related to e-commerce}

\author{Kishore Thapliyal\thanks{tkishore36@yahoo.com} , Anirban Pathak\thanks{anirban.pathak@gmail.com}\\
Jaypee Institute of Information Technology, A-10, Sector-62, Noida,
UP-201307, India}
\maketitle
\begin{abstract}
A set of quantum protocols for online shopping is proposed and analyzed
to establish that it is possible to perform secure online shopping
using different types of quantum resources. Specifically, a single
photon based, a Bell state based and two 3-qubit entangled state based quantum online shopping schemes are proposed. The Bell state based scheme, being a completely orthogonal
state based protocol, is fundamentally different from the earlier proposed
schemes which were based on conjugate coding. One of the 3-qubit entangled state based scheme is build on the principle of entanglement swapping which enables us to accomplish the task without transmission of the message encoded qubits through the channel. Possible ways of generalizing the entangled state based schemes proposed here to the schemes which use multiqubit entangled states is also discussed. Further, all the proposed
protocols are shown to be free from the limitations of the recently proposed protocol
of Huang et al. (Quantum Inf. Process. 14, 2211-2225, 2015) which allows the buyer (Alice) to change her order at a later
time (after initially placing the order and getting it authenticated
by the controller). The proposed schemes are also compared with the existing
schemes using qubit efficiency. 
\end{abstract}

\section{Introduction}

In today's society e-commerce plays a crucial role. Especially, we
often purchase things from online stores. Such a purchase requires
online transaction, and that requires security measures. So far our
online transactions are secured by classical protocols, but it is
well established that the majority of the classical security measures
are vulnerable and they will not remain useful in a post-quantum world
(once a scalable quantum computer is built) \cite{Shor-algo}. Thus,
we need quantum protocols for e-commerce. This is so because security
of any classical cryptographic protocol is based on some assumptions
on the computational power of Eve. In contrast, quantum cryptographic
protocols are unconditionally secure. This fact is known since the
introduction of first quantum key distribution (QKD) protocol in 1984
\cite{bb84}. Since then several quantum protocols have been proposed
for various practical tasks that require security (\cite{bb84,ekert,b92,vaidman-goldenberg,Long and Liu,my book,ping-pong,DLL,lm05,reordering1,switch,With Anindita-pla,With preeti,dsqc-ent swap,dsqcwithteleporta,qd,cdsqc}
and references therein). For example, a set of schemes for QKD \cite{bb84,ekert,b92,vaidman-goldenberg,Long and Liu},
direct secure quantum communication \cite{ping-pong,DLL,dsqcwithteleporta,lm05,reordering1,With Anindita-pla,With preeti,dsqc-ent swap,qd},
and its controlled counterpart-controlled deterministic secure quantum
communication (CDSQC) \cite{switch,cdsqc}, have been proposed in
the recent past (see \cite{my book} for a review). A recent addition
to this long list is quantum e-commerce \cite{e-payment1,e-payment,interbank,interbank-cr-an,online-shop,chou}.
Specifically, in 2014, a three-party quantum protocol for online shopping
was proposed by Chou et al. \cite{chou}. In this protocol, Alice
is a buyer, Bob is a merchant (say a representative of Walmart, Big
Bazaar or any other departmental stores which sales goods through
e-commerce corporation like eBay, Flipkart or Amazon) and Charlie
is a controller who may be considered as a representative of VISA
or Master card or a representative of an e-commerce corporation like,
eBay, Flipkart or Amazon. Chou et al.'s scheme allows Alice to buy
a product from Bob in a secure manner. However, in Chou et al.'s protocol
Charlie can obtain the information encoded by Alice (i.e., which product
she wishes to buy). This limitation of Chou et al.'s protocol \cite{chou}
was noted by Huang et al. \cite{online-shop}, and in 2015, they proposed
an improved protocol for quantum online shopping which is free from
the limitation of Chou et al.'s scheme. More recently, a semiquantum
scheme for quantum online shopping has also been proposed by us \cite{semi}.
Prior to these relatively new protocols, a set of protocols were introduced
for quantum online shopping \cite{e-payment,e-payment1,interbank,interbank-cr-an}.
Specifically, e-payment systems were introduced using quantum group
signature \cite{e-payment1}, quantum blind and group signature \cite{e-payment},
quantum proxy blind signature \cite{interbank}, etc., and they were
critically analyzed. For example, cryptanalysis of inter-bank e-payment
protocol introduced in Ref. \cite{interbank} was performed in \cite{interbank-cr-an}.
Further, schemes based on blind signature have also been designed
using quantum teleportation \cite{proxy-1,blind}. It may be noted
that Chou et al.'s quantum-communication-based protocol \cite{chou}
and its improved version proposed by Huang et al. \cite{online-shop}
are free from the limitations of early protocols \cite{e-payment,e-payment1,interbank,interbank-cr-an}
of quantum online shopping. Thus, we would concentrate on the possible
improvement of Chou et al.'s protocol \cite{chou} and Huang et al.'s
protocol \cite{online-shop}. Further, we aim to provide a large number
of alternative paths that may be used to realize quantum online shopping.
These alternatives would provide choices to the experimentalists interested
in implementing schemes of quantum online shopping, based on the quantum
resources available with them and the noise present in the channel.

In what follows, we will show that the above mentioned schemes of
quantum online shopping \cite{e-payment1,e-payment,interbank,interbank-cr-an,online-shop,chou}
are essentially modified schemes of CDSQC, thus, the protocols of
CDSQC that are introduced in the recent papers of some of the present
authors \cite{cdsqc,semi} can be modified suitably to develop quantum
protocols for online shopping. Further, it will be shown that in Huang
et al.'s protocol the buyer (Alice) can change her order at a later
time (after initially placing the order and getting it authenticated
by the controller). This undesirable feature can be removed by using
our schemes of CDSQC-based online shopping.

The remaining part of the paper is organized as follows. In Section
\ref{sec:Exisiting-protocols-for}, we briefly introduce the existing
schemes of quantum online shopping. Thereafter, in Section \ref{sec:New-protocols-for},
we propose four new protocols for quantum e-commerce
and a few other alternatives to perform the task. Subsequently, in
the same section, we also describe a set of other cryptographic tasks
related to e-commerce which can be realized by slightly modifying
the protocols proposed here. We further discuss the security and qubit
efficiency of the proposed schemes in Sections \ref{sec:Security-of-the}
and \ref{sec:Qubit-efficiency-of}, respectively\@. Finally, we conclude
the paper in Section \ref{sec:Conclusions}.

\section{Existing protocols for quantum online shopping\label{sec:Exisiting-protocols-for}}

Before we briefly describe the existing protocols of quantum online
shopping it will be apt to note a few background information. To begin
with let us note that the security of all the existing protocols of
secure quantum communication has been achieved in
two alternative ways which may be referred to as BB84 subroutine and
GV subroutine. Specifically, an additional set of $n$ qubits are
prepared as verification qubits and inserted randomly in the set of
$n$ travel qubits. The verification qubits are referred to as decoy
qubits and they are used later to check whether any eavesdropping
attempt has been made. The presence of Eve could be inferred from
the error rate computed on these verification qubits. The two alternative
subroutines differ in the types of decoy qubits used and the principle
of security in each case. Specifically, in BB84 subroutine, the security
comes in a manner analogous to the BB84 QKD protocol \cite{bb84}.
Specifically, it comes from non-commutativity and nocloning theorem,
where the decoy qubits are prepared randomly in $\left\{ |0\rangle,\,|1\rangle\right\} $
or $\left\{ |+\rangle,\,|-\rangle\right\} $ basis. The sender and
receiver compute the error rates from the qubits prepared by the sender
and measured by the receiver in the same basis as any eavesdropping
attempt would have led to mismatch. In fact, the attempt made by Eve
to learn the inaccessible information would cause disturbance in an
incompatible observable \cite{New-ref-1}, which is maximal for mutually
unbiased bases \cite{New ref 2}.

In contrast, the sender uses $\frac{n}{2}$ copies of a Bell state
(or equivalently an adequate number of mutliqubit entangled state)
and inserts them randomly in the string of travel qubits in the GV
subroutine. This causes geographical separation between the entangled
qubits (ensured here through application of permutation of particles
(PoP)) and Eve is unaware of the particle pairs, thus may tend to
choose wrong particle pairs to perform Bell measurement leading to
entanglement swapping. Thus, at the receiver's end when correct positions
of the entangled states are available, he can detect the signatures
of eavesdropping in the form of entanglement swapping caused due to
it \cite{With preeti,With chitra IJQI}. An appropriate use of PoP
technique makes the measurement basis unavailable to Eve and security
comes from the orthogonal states only. The name GV subroutine originates
from an orthogonal state based QKD scheme introduced by Goldenberg
and Vaidman, where temporal separation between two
localized wave-packets (whose orthogonal superpositions were used
to send 0 and 1) were used to achieve the security \cite{vaidman-goldenberg}.

Interestingly, the set of all entangled state based protocols which
use BB84 subroutine can be easily transformed to corresponding completely
orthogonal state based schemes by replacing the BB84 subroutine with
GV subroutine (see \cite{With preeti,my book} for detail). 

Also note that all the protocols described in this section and the
next section assume that the buyer (Alice) and merchant (Bob) are
registered members of eBay (eBay is an e-commerce corporation, but
it can be equivalently viewed as an online portal or a bank, too),
and eBay (Charlie) is capable to authenticate the identities of Alice
and Bob while communicating with them. Further, in all the protocols
described below, Alice sends a classical information $M$ via quantum
means to Bob. Here, $M$ is the shopping information of Alice which
includes her customer id, items to be purchased (item numbers), quantity
of each item to be ordered, etc. 

To begin with let us briefly describe Chou et al.'s protocol \cite{chou},
which we refer here as CLZ protocol.

\subsection{Chou, Lin, Zeng (CLZ) protocol}
\begin{description}
\item [{CLZ~1:}] Alice informs Charlie that she wants to purchase something
online. After receiving this information, Charlie prepares and sends
her a sequence of $2n$ qubits that is randomly prepared in $\left\{ |0\rangle,|1\rangle,|+\rangle,|-\rangle\right\} $.
However, Charlie does not disclose which qubit is prepared in which
basis. \\
Out of these $2n$ qubits, $n$ will be used as decoy/verification
qubits which will be used for eavesdropping check. \\
In the original CLZ protocol, Charlie used to send $n+\delta$ qubits,
out of which $\delta$ were verification qubits, but for unconditional
security we need to check half of the received qubits \footnote{In Ref. \cite{nielsen}, it is shown that a random test of half of
the qubits provide an upper bound on the errors in the rest of the
transmitted qubits.}. This is why we use $\delta=n$ in this paper.
\item [{CLZ~2:}] Alice randomly selects $n$ of the $2n$ qubits received
by her and in collaboration with Charlie applies BB84 subroutine on
those $n$ qubits. If the computed error rate is found to be lesser
than the tolerable limit they continue to the next step, otherwise
they restart the protocol. \\
After the eavesdropping check is performed using BB84 subroutine,
the qubits used for the same are discarded, and Alice is left with
$n$ qubits which she uses as message qubits in the next step.
\item [{CLZ~3:}] Alice encodes her shopping information $(M)$ on the
$n$ qubits of her possession using following rule: to encode $0\,(1)$
she does nothing (applies $iY$ operator). Subsequently, she randomly
inserts $n$ decoy qubits prepared at random in $\left\{ |0\rangle,|1\rangle,|+\rangle,|-\rangle\right\} $
into the message encoded sequence and sends that to Bob. \\
The encoding operation here is the same as that used in LM05 protocol
\cite{lm05} of QSDC. 
\item [{CLZ~4:}] After receiving an authenticated acknowledgment of the
receipt of $2n$ qubits from Bob, Alice discloses the positions of
$n$ decoy qubits, and Alice and Bob apply BB84 subroutine on the
decoy qubits to check eavesdropping. If no eavesdropping is found
they go to the next step, otherwise they restart the protocol.
\item [{CLZ~5:}] Bob asks Charlie, for the initial states of the $n$
message qubits available with him, and Charlie provides that information.
With the encoded qubits and their initial states, merchant Bob can
now deduce the shopping information of the customer.
\end{description}

\subsection{Huang, Yang, Jia (HYJ) protocol}

Huang et al. \cite{online-shop} showed that in CLZ 3, when Alice
sends a message encoded sequence to Bob, Charlie can capture all the
qubits and replace them by a fake sequence of $2n$ qubits. Later,
when in CLZ 4, Alice discloses the positions of decoy qubits, Charlie
discards them from the captured sequence and measures rest using the
basis in which he had initially prepared the qubits. Eavesdropping
check will definitely reveal this, but by then Charlie will have all
the information encoded by Alice although this information was not
intended for him. For instance, assume that a country wishes to buy
some items for defense (say, weapons) from a multinational company,
which would prefer to keep the details of the deal secret as its policy,
and the buyer would not like to reveal this information (specifically,
which items he is going to buy and the quantities of each item to
be purchased) to a third party helping in making the payments. This
limitation of CLZ protocol was circumvented in the HYJ protocol, which
may be viewed as an improved version of the CLZ protocol. HYJ protocol
may be described as follows:
\begin{description}
\item [{HYJ~1:}] Same as CLZ 1.
\item [{HYJ~2:}] Same as CLZ 2.
\item [{HYJ~3:}] Same as in CLZ 3 with a difference that Alice also prepares
a random key $K$, and instead of $M$ she sends $M^{\prime}=K\oplus M$
to Bob and keeps $K$ secret.
\item [{HYJ~4:}] Same as CLZ 4.
\item [{HYJ~5(a):}] Alice announces $K$, and Bob uses that to obtain
$M=K\oplus M^{\prime}.$
\item [{HYJ~5(b):}] Same as CLZ 5.
\end{description}
As $K$ is unknown to Charlie till eavesdropping is checked in HYJ
4, Charlie's strategy of sending fake sequence of qubits to Bob will
not work here. Specifically, Charlie's eavesdropping will be detected
in HYJ 4, and $K$ is announced in HYJ 5 if and only if no trace of
eavesdropping is detected in HYJ 4. Thus, HYJ scheme is free from
the limitation of the CLZ protocol. However, there exists a limitation
of HYJ protocol as we have already mentioned. As $K$ is not known
to Bob, Alice has freedom to change his order till HYJ 5(a) by disclosing
a different key $K^{\prime}.$ Let us explain this with a specific
example. Consider that if Alice wants to buy a TV she has to send
$M_{1}=100101$; and if she wants to buy a refrigerator then she has
to send $M_{2}=111100$. In HYJ 3, Alice uses $K=010010$ as key to
yield $M^{\prime}=K\oplus M_{1}=010010\oplus100101=110111,$ but at
the time of disclosure of the key she changes her mind and announces
$K^{\prime}=001011$ as her key. As a consequence, Bob will decode
the message as $K^{\prime}\oplus M^{\prime}=001011\oplus110111=111100=M_{2}$.
It is possible to cease this freedom of Alice to change order till
the end. However, to do so we have to ensure that Alice is not able
to change the key after HYJ 3. It is possible if a copy of the key
is already available with Bob. Thus, Alice and Bob need to implement
a protocol of QKD, quantum key agreement, direct secure quantum communication
first to create or distribute a key and subsequently use that key
for encryption. This would restrict Alice from changing her order
at the last moment.

One may argue that this freedom to choose the merchandise for Alice
cannot affect the task intended. However, suppose that possibility
of placing an order is allowed only for the goods available in the
store. In that case, Alice may change her order to buy anything that
may be available later. Further, on numerous such occasions (as in
online limited offer sale, which start on a predecided time and the
orders made before and after the start of the sale must be treated
independently), this freedom of Alice is not desired. In what follows,
we describe a few new schemes for quantum online shopping which are
free from the above mentioned limitation of HYJ protocol.

\section{New protocols for quantum online shopping\label{sec:New-protocols-for}}

In this section, we report four different protocols for quantum online
shopping. All of them are essentially modified schemes for CDSQC.

\subsection{Protocol 1: PoP based quantum online shopping using single photons}

Firstly, we show a simple minded PoP based scheme which is equivalent
to HYJ protocol. The protocol is described as follows (remaining steps
are the same as that in HYJ protocol).
\begin{description}
\item [{PoP~3:}] Same as in CLZ 3 with a difference that Alice applies
a permutation operator $\Pi_{n}$ on her message encoded sequence
before random insertion of the decoy qubits, but keeps the actual
sequence secret. 
\item [{PoP~5(a):}] Alice announces $\Pi_{n}$ and Bob uses that to obtain
$M.$
\end{description}

\subsection{Protocol 2: Orthogonal state based quantum online shopping using
Bell states}

Our orthogonal state based quantum online shopping protocol can be
described in following steps:
\begin{description}
\item [{OSB~1:}] Charlie prepares $n$ Bell states $|\psi^{+}\rangle^{\otimes n}$
with $n\geq2$, where $|\psi^{\pm}\rangle=\frac{|00\rangle\pm|11\rangle}{\sqrt{2}}$
and $|\phi^{\pm}\rangle=\frac{|01\rangle\pm|10\rangle}{\sqrt{2}}$.
He prepares two ordered sequences from the Bell states as follows: 
\end{description}
\begin{enumerate}
\item A sequence of all the first qubits of the Bell states: $P_{A}=\left[p_{1}\left(t_{A}\right),p_{2}\left(t_{A}\right),...,p_{n}\left(t_{A}\right)\right]$, 
\item A sequence of all the second qubits of the Bell states: $P_{B}=[p_{1}(t_{B}),p_{2}(t_{B}),...,p_{n}(t_{B})]$,\\
where the subscripts $1,2,\cdots,n$ denote the order of a particle
pair $p_{i}=\{t_{A}^{i},t_{B}^{i}\},$ which is in the Bell state.
\end{enumerate}
\begin{description}
\item [{OSB~2:}] Charlie applies an $n$-qubit permutation operator $\Pi_{n}$
on $P_{B}$ to create a new sequence as $P_{B}^{\prime}=\Pi_{n}P_{B}$. Charlie withholds the information
of the actual order ($\Pi_{n}$) to restrict Bob from decoding Alice's
message prior his permission to do so. Therefore, Bob need not bother
about Alice's choice of merchandise before Charlie informs that payment
has been made by her. \\
Without the knowledge of the permutation operator $\Pi_{n}$, a potential
eavesdropper will also be ignorant about the particle pairs in the
Alice's and Bob's strings $P_{A}$ and $P_{B}^{\prime}$, respectively. 
\item [{OSB~3:}] Charlie subsequently prepares $2n$ decoy qubits as $|\psi^{+}\rangle^{\otimes n}$
and randomly inserts the first (last) $n$ Bell states as decoy qubits
in $P_{A}$ ($P_{B}^{\prime}$) to yield a larger sequence $P_{A}^{\prime\prime}\,(P_{B}^{\prime\prime})$
having $2n$ qubits. \\
It would be relevant to mention that the choice of Bell states as
decoy qubits (i.e., GV subroutine) is not unique here. The same task
can also be achieved using BB84 subroutine\footnote{The BB84 and GV subroutines are shown to be equivalent in the ideal
conditions, while over noisy channels this equivalence does not hold
anymore \cite{decoy}.} or using other entangled states. As we are restricting ourselves to
a completely orthogonal state based online shopping scheme we are
discussing only random insertion of Bell pairs in the sequence of
message qubits. Finally, Charlie sends $P_{A}^{\prime\prime}$ and
$P_{B}^{\prime\prime}$ to Alice and Bob, respectively.
\item [{OSB~4:}] Charlie discloses the positions of the decoy qubits and
partner pairs in Bell states after receiving the authenticated acknowledgment
of the receipt of the qubits from Alice and Bob. Alice and Bob apply
GV subroutine to check error rate and if the computed error rate is
found lower than the tolerable error limit, they go to the next step.
Otherwise, they return back to OSB1.
\item [{OSB~5:}] Alice can now encode her shopping information $M$ by
performing suitable Pauli operation. It is predecided that $I,\,X,\,iY$,
and $Z$ Pauli operations will be used to encode 00, 01, 10, and 11,
respectively. Subsequently, Alice concatenates $n$ decoy qubits (with
a prior intention to use GV subroutine for eavesdropping checking)
in the message encoded qubits. Finally, Alice sends the enlarged sequence
$P_{A}^{\prime}$ to Bob after applying the permutation operator $\Pi_{2n}^{\prime}$.
\\
Though Bob has now access to both $P_{A}^{\prime}$ and $P_{B}^{\prime}$,
he will not be able to find out which particle is entangled with which
particle and decode Alice's message. Thus, he needs Charlie's (Alice's)
disclosure of $\Pi_{n}$ $\left(\Pi_{2n}^{\prime}\right)$ before
decoding the message. Alice's permutation operation provides security
against the Charlie's aforementioned participant attack \cite{online-shop}.
\item [{OSB~6:}] Alice discloses $\Pi_{n}^{\prime}$ corresponding to decoy qubits. Alice and Bob perform GV subroutine
on the decoy qubits Alice has inserted. If the errors are below tolerable
limit they proceed, otherwise they abort the protocol.
\item [{OSB~7:}] Charlie discloses $\Pi_{n}$ and Alice announces $\Pi_{n}^{\prime}$,
which allow Bob to decode Alice's message.
\item [{OSB~8:}] Since the initial Bell states and exact sequence are
known, Bob measures the initially entangled partner pairs in the Bell
basis and using the outcomes of his measurement, he decodes the order
information sent by Alice. 
\end{description}
This quantum online shopping scheme is using the idea of quantum cryptographic
switch discussed in Refs. \cite{cdsqc,switch,crypt-switch}. 

\subsection{Protocol 3: Quantum online shopping using entanglement swapping \label{subsec:Alternative-2:-CDSQC}}

We will now introduce an entanglement swapping based quantum online
shopping scheme, inspired by the direct secure quantum communication
\cite{dsqc-ent swap} and CDSQC \cite{cdsqc} protocols based on entanglement
swapping where communication is performed without actually transmitting
the message encoded qubits. Specifically, our entanglement swapping
based scheme works as follows.
\begin{description}
\item [{ESB~1:}] Charlie prepares $n$ copies of a three qubit GHZ-like
entangled state 
\begin{equation}
|\psi\rangle=\frac{1}{\sqrt{2}}\left(|\psi_{1}\rangle_{12}|a\rangle_{3}\pm|\psi_{2}\rangle_{12}|b\rangle_{3}\right),\label{eq:3-qubit state}
\end{equation}
where $|\psi_{i}\rangle$s are the Bell states such that $|\psi_{1}\rangle\neq|\psi_{2}\rangle$,
and the single qubit states $|a\rangle$ and $|b\rangle$ are orthogonal
to each other, i.e., $\langle a|b\rangle=\delta_{a,b}$. This restriction
ensures that qubit 3 remains appropriately entangled
with the remaining 2 qubits. \\
For instance, without loss of generality, we can assume that Charlie
prepares $|\psi\rangle=\frac{1}{\sqrt{2}}\left(|\psi^{+}\rangle_{12}|0\rangle_{3}+|\psi^{-}\rangle_{12}|1\rangle_{3}\right).$
\item [{ESB~2:}] Charlie prepares three strings $P_{A1}$, $P_{A2}$,
and $P_{B}$ of qubits 1, 2, and 3 of all $n$ GHZ-like entangled
state, respectively. He further performs a permutation operator $\Pi_{n}$
on $P_{B}$ to obtain $P_{B}^{\prime}$. This permutation operator
ensures Charlie's control over the communication between Alice and
Bob.
\item [{ESB~3:}] Subsequently, Charlie inserts $n$ decoy qubits randomly
in all three strings to obtain enlarged strings $P_{A1}^{\prime}$,
$P_{A2}^{\prime}$, and $P_{B}^{\prime\prime}$. The choice of decoy
qubits depends upon type of subroutine predecided by the legitimate
parties to be performed for eavesdropping checking. Thereafter, he
sends $P_{A1}^{\prime}$ and $P_{A2}^{\prime}$ $\left(P_{B}^{\prime\prime}\right)$
to Alice (Bob). 
\item [{ESB~4:}] Same as OSB 4 but the choice of eavesdropping checking
subroutine is arbitrary.
\item [{ESB~5:}] Alice prepares $n$ copies of $|\psi^{+}\rangle_{A_{1}A_{2}}$
to encode her secret information of items to be purchased. Specifically,
she applies a $Z$ gate on one of the qubits of the Bell state to
encode 1 and does nothing to send 0. Therefore, the composite state
Alice and Bob hold is $|\psi^{\prime}\rangle=\frac{1}{\sqrt{2}}\left(|\psi^{\pm}\rangle_{A_{1}A_{2}}|\psi^{+}\rangle_{12}|0\rangle_{3}+|\psi^{\pm}\rangle_{A_{1}A_{2}}|\psi^{-}\rangle_{12}|1\rangle_{3}\right).$
\item [{ESB~6:}] Alice measures qubits $A_{1}$ and 1 as well as $A_{2}$
and 2 in Bell basis, while Bob can measure his qubits in the computational
basis. Subsequently, she announces her measurement outcomes, which
should reveal her message to Bob.\\
To understand this point we can write the state before measurement
as 
\begin{equation}
\begin{array}{lcl}
|\psi^{\prime}\rangle_{m} & = & \frac{1}{2\sqrt{2}}\left(\left\{ |\psi^{+}\rangle_{A_{1}1}|\psi^{+}\rangle_{A_{2}2}+|\psi^{-}\rangle_{A_{1}1}|\psi^{-}\rangle_{A_{2}2}\pm|\phi^{+}\rangle_{A_{1}1}|\phi^{+}\rangle_{A_{2}2}\pm|\phi^{-}\rangle_{A_{1}1}|\phi^{-}\rangle_{A_{2}2}\right\} |m\rangle_{3}\right.\\
 & + & \left.\left\{ |\psi^{+}\rangle_{A_{1}1}|\psi^{-}\rangle_{A_{2}2}+|\psi^{-}\rangle_{A_{1}1}|\psi^{+}\rangle_{A_{2}2}\mp|\phi^{+}\rangle_{A_{1}1}|\phi^{-}\rangle_{A_{2}2}\mp|\phi^{-}\rangle_{A_{1}1}|\phi^{+}\rangle_{A_{2}2}\right\} |\overline{m}\rangle_{3}\right),
\end{array}\label{eq:msg}
\end{equation}
where $m$ corresponds to the message bit encoded by Alice and the
upper (lower) sign in the right-hand side of the equation corresponds
to $m=0$ (1). Thus, if Alice announces both her measurements resulted
in the same (different) Bell state(s) then Bob's measurement outcome
will be same as (different from) the bit encoded by Alice, i.e., $m$
$\left(\overline{m}\right)$.
\item [{ESB~7:}] Same as OSB 7.
\end{description}
Note that the Bell state prepared by Alice to encode her message was
measured by her only in ESB 6. Therefore, message encoded qubits actually
do not travel through the channel accessible to Eve at all. This further
reduces the number of times eavesdropping checking is to be performed
as the rounds of transmission of qubits is reduced. In view of some
of our recent results \cite{Vishal,NM,AQD,QC}, which show that the
performance of a quantum cryptographic scheme (in presence of noise)
decays with an increase in the number of rounds quantum communication
is involved, we can predict that this scheme would be relatively more
robust against channel noise. 

\subsection{Protocol 4: Quantum online shopping using dense coding \label{subsec:Alternative-3:-CDSQC}}

The GHZ-like entangled state discussed in the previous protocol can
also be used to send message in a secure manner without relying on
entanglement swapping.
\begin{description}
\item [{DCB~1:}] Same as ESB 1.
\item [{DCB~2:}] Charlie prepares three strings $P_{A}$, $P_{B}$, and
$P_{C}$ of qubits 1, 2, and 3 of all GHZ-like entangled states, respectively.
Here, Charlie need not perform a permutation operator as he can also
ensure his control over the communication by keeping the third qubit
with himself.
\item [{DCB~3:}] Subsequently, Charlie inserts $n$ decoy qubits randomly
in both strings $P_{A}$ and $P_{B}$ to obtain enlarged strings $P_{A}^{\prime}$
and $P_{B}^{\prime}$, respectively. The choice of decoy qubits depends
upon type of subroutine predecided by the legitimate parties to be
performed for eavesdropping checking. Thereafter, he sends $P_{A}^{\prime}$
and $P_{B}^{\prime}$ to Alice and Bob, respectively. 
\item [{DCB~4:}] Same as ESB 4.
\item [{DCB~5:}] Same as OSB 5, but Bob requires Charlie's measurement
results for the string $P_{C}$. Also the eavesdropping checking subroutine
is arbitrary.
\item [{DCB~6:}] Same as OSB 6.
\item [{DCB~7:}] Same as OSB 7, while Charlie announces the result of
measurement performed on the third qubit in $\left\{ |a\rangle,|b\rangle\right\} $
basis instead of the permutation operator.
\item [{DCB~8:}] Same as OSB 8.
\end{description}
So far we have discussed the possibility of performing quantum online
shopping task with the help of single photons (in Protocol 1), Bell
states (in Protocol 2), and three-qubit GHZ-like state (in Protocols
3 and 4). However, it is not limited to this small set of states.
There exist several alternative approaches through which efficient
online shopping schemes can be designed using a large class of states.
Here, we briefly mention a generalized approach to all these alternative
paths.

\subsection{Various alternative ways to perform quantum online shopping using
multi-qubit entangled states \label{subsec:Alternative-1:-CDSQC}}

The task in hand can be accomplished using other multi-qubit entangled
states, too. Specifically, in a densecoding based direct communication
scheme \cite{qd,AQD,cdsqc}, Alice (Bob) possesses $p\leq\frac{N}{2}$
$\left(N-p\right)$ qubits out of total $N$ qubits of a $N$-qubit
entangled state (such as $W$ state, GHZ state, GHZ-like state, $Q_{4}$
state, $Q_{5}$ state, cluster state, $|\Omega\rangle$ state, Brown
state). Alice encodes her message using a suitable set of unitary
operations and sends the qubits to Bob, who measures all $N$ qubits
in the basis they were prepared.

In the densecoding based online shopping scheme (analogous to Protocol
2), Charlie can randomly prepare one of the above mentioned multi-qubit
states and sends $p$ qubit to Alice while $N-p$ qubits to Bob in
a secure manner. He permutes Bob's qubits to maintain his control
power. Thereafter, Alice and Bob can perform the task under Charlie's
supervision.

Along the line of Protocol 3, a quantum online shopping scheme using
entanglement swapping that can transmit an $s$-bit message can be
designed using the quantum states of the form 
\begin{equation}
|\psi\rangle=\frac{1}{\sqrt{2^{s}}}\sum_{i=1}^{2^{s}}|e_{i}\rangle|f_{i}\rangle,\label{eq:state of interest}
\end{equation}
where $|e_{i}\rangle$ is an N-qubit maximally entangled state (as
$W$ state, GHZ state, GHZ-like state, $Q_{4}$ state, $Q_{5}$ state,
cluster state, $|\Omega\rangle$ state, Brown state). Specifically,
$\left\{ |e_{i}\rangle\right\} $ is a basis set with maximally entangled
basis vectors in $C^{2^{N}}:\,N\geq s$, while $\left\{ |f_{i}\rangle\right\} $
is a basis set in $C^{2^{l}}:l\geq s\geq1$ which may be separable.
Thus, $|\psi\rangle$ (an $N+l$ qubit entangled state) is prepared
by Charlie, who shares the string of first $N$ qubits with Alice
and that of the last $l$ qubits with Bob in a secure manner. He had
to perform a permutation operator on Bob string to maintain his control
over shopping scheme. Thus, Alice can send her message by preparing
extra $m$-qubit entangled state in $\left\{ |e_{i}\rangle\right\} $
basis and measure her qubits in such a way that message is transmitted
through entanglement swapping \cite{dsqc-ent swap}. 

A quantum online shopping scheme without entanglement swapping can
also be performed using quantum channel of the form in Eq. (\ref{eq:state of interest}).
Specifically, Charlie keeps the last $l$ qubits with himself and
sends $p\leq\frac{N}{2}$ $\left(N-p\right)$ qubits from the remaining
qubits to Alice (Bob) in a secure manner. Using densecoding Alice
encodes her message and sends $p$ qubits in a secure manner to Bob,
who requires Charlie to inform him the measurement outcome of the
$l$ qubits to decode Alice's message. Therefore, we have discussed
a set of possible mechanism to obtain quantum online shopping schemes
using a large set of quantum states.

Once Alice and Bob share an entangled state prepared by Charlie, she
can teleport her choice of item to be purchased to Bob \cite{dsqcwithteleporta}.
Analogous to all the schemes discussed so far he will need Charlie's
assistance to reconstruct the details of Alice's order.

\subsection{Other tasks related to e-commerce \label{sec:Other-tasks-related}}

With an escalated interest on quantum internet \cite{QInternet,QInternet2}
the feasibility of implementation of quantum solutions for e-commerce,
voting \cite{Voting1}, sealed-bid auction \cite{Our-auction}, etc.,
has also enhanced. Therefore, here we list a set of tasks those can
be performed using modified forms of the proposed schemes. So far
we were considering the quantum online shopping where the merchant
delivers the order to buyer once payment has been made and confirmed
by his bank. We can consider nowadays with the advent of online availability
of soft copies of books, magazines, audio, and video which would require
a bidirectional communication. Specifically, a quantum bidirectional
online shopping scheme based on controlled quantum dialogue \cite{crypt-switch,NM}
can be performed where Alice sends information of her order to Bob
under the supervision of Charlie, while Bob sends some sample files
to Alice. Depending upon the quality of sample file, Alice may choose
to cancel her order for which her payment will be refunded. Protocols
described here can be easily modified to design many schemes for secure
online computation and communication. For example, we may think of
online examinations (where a central body (Charlie) would authorize
an examiner (Bob) either to evaluated the answer sheet of the candidate
Alice or to send him the question paper/evaluated answer sheet), participation
in webinars (where the convener (Charlie) would check whether the
participant Alice is authorized (paid registration fees) to listen
the talk of Bob, viewing of a particular channel in the TV set of
a particular customer (where the service provider (Charlie) would
check whether the customer Alice is authorized to see the Channel
broadcasted by Bob. Multicontroller versions of all such schemes can
also be designed \cite{chou}. We are not extending this list here,
but numerous such situations exist where the current protocol or its
simple variants will be useful.

\section{Security of the proposed protocols \label{sec:Security-of-the}}

As the set of protocols proposed by us are variants of CDSQC schemes,
in analogy of the CLZ \cite{chou} or HYJ \cite{online-shop} protocols,
the security of the schemes can be established along the same line.
However, the attack designed in Ref. \cite{online-shop} and loophole
pointed out by us here are not applicable on the proposed schemes.
In what follows, we can categorize the security against a set of attacks
in the outsider's and participant's attacks. 

\subsection{Security against outsider's attack}

Here, we will establish the security of the proposed schemes against
Eve's individual attacks.
\begin{enumerate}
\item \textbf{Intercept-and-resend attack:} Eve may choose to replace the
qubits sent by Charlie (or a sender in general) by freshly prepared
qubits and send it to the receiver. Using this attack, Eve would succeed
to get Alice's message if she intercepts the encoded qubits sent from
Alice to Bob exploiting information of the initially prepared state.
To circumvent this attack in Protocol 1, BB84 subroutine is performed
for eavesdropping checking. Specifically, when Charlie and Alice (or
Bob) compute error rates using the randomly inserted decoy qubits
prepared in two mutually unbiased bases, and from which they can proceed
with (discard) the protocol if the error rate is below (above) threshold.
If Eve measures the intercepted qubits either in the computational
or diagonal basis before sending the freshly prepared qubit in the
measurement outcome, she would induce disturbance for the wrong choice
of basis. Specifically, for $n$ travel qubits, eavesdropping checking
is performed on $\frac{n}{2}$ decoy qubits. Suppose Eve measures
$m$ qubits (which will have both decoy and message qubits). Without
loss of generality, we can assume that an equal number of decoy and
message qubits are measured, i.e., $\frac{m}{2}$ decoy qubits are
measured out of the total number of $\frac{n}{2}$ decoy qubits. As
the error rate is calculated using decoy qubits only, the fraction
of the intercepted decoy qubits is $f=\frac{m/2}{n/2}=\frac{m}{n}$.
Thus, the mutual information between the sender (Charlie) and Eve
is $I(C:E)=f/2$ as she (Eve) can choose the correct basis half of
the time. In the rest half of the time, when she chooses the wrong
basis, she would prepare states in the wrong basis resulting in the
wrong measurement outcome at the receiver's end half of the time $e=\frac{f/2}{2}=\frac{f}{4}$.
Thus, $I(A:B)=(1-H[\frac{f}{4}])$, where $H\left[u\right]$ is the
Shannon binary entropy. A quantum cryptographic scheme works until
$I(A:B)\geq I(A:E)$, which results in the present case as tolerable
attack fraction $f\cong0.68$ and corresponding error rate as $17\%$
(\cite{QC,AQD} and references therein). Note that Eve's success probability
for each attacked qubit is $\frac{3}{4}$, which would become very
small for higher values of $m$ as it would approach $\left(\frac{3}{4}\right)^{m}$.
\item \textbf{Entangle-and-measure attack:} Instead of intercepting and
measuring the transmitting qubit, Eve may choose to entangle her qubit
(initially prepared in $\alpha\left|0\right\rangle +\beta\left|1\right\rangle $)
with the travel qubits accessible to her. To obtain the secret she
will measure her ancillae qubits at a later stage. However, during security
checking, decoy states randomly prepared in $\left|0\right\rangle ,$
$\left|1\right\rangle ,$ $\left|+\right\rangle $, and $\left|-\right\rangle $
are measured in the computational and diagonal basis, which would
result the wrong result with probability $\left|\beta\right|^{2}$
if she attacks $\left|0\right\rangle $ or $\left|1\right\rangle $
decoy state, while the states remain separable for the rest of the
decoy qubits. Thus, total probability of detection of Eve in this
attack is $\frac{\left|\beta\right|^{2}}{2}$ assuming all four decoy
states are equally probable \cite{QC,AQD}.
\item \textbf{Correlation-elicitation attack:} In this attack, Eve may exploit
availability of some of the qubits of the entangled quantum channel
more than once to extract encoded message \cite{Cor-eli}. Specifically,
Eve can check the parity of Bell states using two CNOT if both the
qubits are accessible to her. In the proposed schemes, use of decoy
qubits leaves signature of eavesdropping to be detected while checking
subroutine.
\item \textbf{Impersonation attack or Man-in-the-middle attack:} Eve may
try to play as the receiver (sender) to the sender (receiver). This
attack can be prevented by using authentication of the identities
of the sender and receiver before quantum communication \cite{authentication1,authentication2}.
The receiver should further inform the sender about the receipt of
the transmitted qubits over such an authenticated classical channel.
\item \textbf{Disturbance attack or modification attack:} Eve may also attempt
some denial of service attacks, where she does not intend to extract
information but to misguide the legitimate parties only. She can disturb
the content of the message by changing the order of qubits or applying
random unitary operations on the encoded qubits. Note that this will
not reveal any information to Eve but this attack will also be revealed
during eavesdropping checking performed to ensure secure transmission
of the qubits (\cite{AQD} and references therein).\\ 
A special case which requires attention is the orthogonal state based quantum online shopping scheme which is described above as Protocol 2. In Protocol 2, if Eve applies a single qubit operation (say $X$ gate) on all the transmitted qubits, it will not change the decoy states being two qubit Bell states, while the order information will be changed completely. Applicability of this attack is not restricted to the Protocol 2 of e-commerce, it's in fact applicable to many schemes whose security is ensured by using the GV subroutine. This is not a serious attack as it does not reveal any information to Eve, but it has a serious impact on e-commerce as it can be used to change the order. Interestingly, it's possible to  circumvent this attack. To do so, Alice can send some redundant qubits and compare their measurement outcomes with Bob. 
\item \textbf{Trojan-horse attack:} Eve may design attacks based on the
implementation of the scheme \cite{Rev-cryp,Troj-hor,Troj-hor1}.
Though a quantum cryptographic scheme can not be proved secure against
trojan-horse attack using principles of quantum mechanics, various
technical measures have been discussed in the recent past to circumvent
these attacks \cite{Rev-cryp,Troj-hor,Troj-hor1}. 
\end{enumerate}

\subsection{Security against participant's attack}

A participant attack is more powerful in multiparty quantum communication
scheme as a legitimate user has more access to the useful information
than Eve. Therefore, here we analyze the attacking strategy of each
participant.
\begin{enumerate}
\item Charlie can try to extract the choice of merchandise by Alice as pointed
out in Ref. \cite{online-shop}, where this attack has been circumvented
by the use of an extra key by Alice. However, here we have discussed
that this provide an advantage to Alice, which would not be present
if Alice and Bob share the quantum key used by her using a quantum
key distribution or agreement scheme (\cite{bb84,b92,semi} and references
therein). Here, we have proposed that Alice applies a permutation
operation on the encoded qubits to defend from this attack. \\
Charlie can also exploit the fact that he is authorized to prepare
the state to be used as quantum channel. As the proposed schemes would
remain secure until the controller prepares the desired state, Alice
and Bob can randomly choose to measure a few copies of such states
to check correlations. 
\item Alice can try to cheat by changing the item ordered, without being
detected, even after payment has been made. We have pointed the feasibility
of this attack in the e-commerce schemes proposed in Refs. \cite{online-shop,chou}.
However, the present schemes take care of any such attack by either
using a key which is already shared between Alice and Bob or by application
of a permutation operation by Alice. In the latter case, if Alice
chooses to reveal wrong permutation operator, the information shared
by Charlie regarding the initial state preparation to Bob will fail
to extract order details. Thus, the online shopping task will be aborted
and Alice would not gain any advantage.
\item As mentioned in Ref. \cite{chou}, Bob can try to know the order placed by Alice before
Charlie authorizes him to do so. However, Bob will not be able to decode information regarding Alice's order without Charlie's assistance as he will remain ignorant about the initial state chosen by the supervisor.
\end{enumerate}

\section{Qubit efficiency of the protocols\label{sec:Qubit-efficiency-of}}

The efficiency of a secure quantum communication scheme can be analyzed
using a quantitative measure \cite{defn  of qubit  efficiency} defined
as 
\begin{equation}
\eta=\frac{c}{q+b},\label{eq:eff}
\end{equation}
where $c$ is the number of classical bits transmitted using \textbf{$q$
}number of qubits. In addition, $b$ bits of classical communication
is also required, which does not include that used for eavesdropping
checking.

There is one more parameter, often discussed to analyze the performance
of a cryptographic scheme, which quantifies number of bits transmitted
per qubit in a scheme. Note that this definition does not take into
account the classical communication required in accomplishing certain
task. Thus, we can use another quantitative measure as 

\begin{equation}
\eta_{q}=\frac{c}{q}.\label{eq:eff-q}
\end{equation}

\begin{table}
\begin{centering}
\begin{tabular}{c>{\centering}p{2cm}c}
\hline 
Name of the protocol & $\eta$ & $\eta_{q}$\tabularnewline
\hline 
CLZ \cite{chou}  & $\frac{1}{4}$ & $\frac{1}{3}$\tabularnewline
HYJ \cite{online-shop}  & $\frac{1}{5}$ & $\frac{1}{3}$\tabularnewline
Semiquantum online shopping Protocol 1 \cite{semi}  & $\frac{1}{23}$ & $\frac{1}{21}$\tabularnewline
Semiquantum online shopping Protocol 2 \cite{semi}  & $\frac{1}{18}$ & $\frac{1}{16}$\tabularnewline
Protocol 1 proposed here & $\frac{1}{5}$ & $\frac{1}{3}$\tabularnewline
Protocol 2 proposed here & $\frac{2}{7}$ & $\frac{2}{5}$\tabularnewline
Protocol 3 proposed here & $\frac{1}{8}$ & $\frac{1}{6}$\tabularnewline
Protocol 4 proposed here & $\frac{2}{9}$ & $\frac{1}{3}$\tabularnewline
\end{tabular}
\par\end{centering}
\caption{\label{tab:comparison of efficiency} Comparison of qubit efficiency
of the existing and the proposed protocols.}
\end{table}

The comparative study performed here (summarized in Table \ref{tab:comparison of efficiency}) establishes that the proposed
protocols are much efficient compared to the existing schemes \cite{chou,online-shop,semi}.
Specifically, the qubit efficiency of the improved HYJ protocol is
less than that of CLZ protocol as an extra $n$ bit classical communication
of key by Alice is required in the former protocol. Our Protocol 1,
which uses the same amount of quantum and classical resources as HYJ
scheme, is equally efficient to that. The rest of our protocols use
entangled states. Specifically, Protocol 2 uses Bell states and densecoding
thus becomes most efficient scheme. Note that Protocols 3 and 4 use
GHZ-like states while difference in qubit efficiency comes from the
reason that densecoding cannot be used in Protocol 3, and Protocol
4 requires $n$ bit classical communication. Thus, Protocol 4 is also
more efficient than previously existing HYJ protocol. Protocol 3 may
not appear more efficient but has an intrinsic advantage that message
encoded qubits are never accessible to Eve. This feature is quite
consistent with our earlier observation that requirement of quantum
resources increases for sophisticated and complex quantum cryptographic
tasks \cite{semi,QPC}. In fact, one can clearly see that semiquantum online shopping schemes \cite{semi} have very small values of both qubit efficiency and bits transmitted per qubit used in comparison to the rest of the online scheme in Table \ref{tab:comparison of efficiency}.  

If the number of bits transmitted per qubit used is considered then
our Protocols 1 and 4 are equally efficient as CLZ and HYJ. Protocol
2 (3) is most (least) efficient in the set of quantum online shopping
schemes discussed here. We have not compared the qubit efficiency
of the proposed schemes with more recent schemes \cite{blind,proxy-1}
due to use of quantum teleportation and quantum key distribution in
them to ensure the security of the online banking. Also, the semiquantum
schemes for any cryptographic task are also less efficient than corresponding
fully quantum schemes \cite{semi}.

\section{Conclusions \label{sec:Conclusions}}

With advent of quantum technologies and feasibility of quantum internet
in the near future, various socio-economic problems can be addressed
using quantum solutions. Along this line quantum schemes for voting,
auction and online banking have been introduced in the recent past.
As far as online shopping is concerned, an improvement in the recently
proposed single photon based scheme \cite{chou} has been proposed,
which restricts the bank from accessing the order placed by the buyer
\cite{online-shop}. Here, we have pointed out a loophole in the improved
scheme \cite{online-shop} that the buyer may change the merchandise
even after payment for his order has been made. 

We further propose solution to such a participant attack by using
a previously shared key instead of random key by the buyer or a permutation
operator. Here, we have given a set of such schemes using different
quantum channels based on various CDSQC schemes. Specifically, a single
photon based improved scheme (Protocol 1), a cryptographic switch
based scheme using Bell state (Protocol 2), and two 3-qubit state
based schemes using entanglement swapping (Protocol 3) and densecoding
(Protocol 4) are proposed. The entangled state based schemes are further
shown generally implementable using a large set of schemes. Thus, the proposed schemes provide numerous possible
ways to experimentally implement the quantum online shopping scheme. 

From the set of proposed quantum online shopping schemes the single
photon based proposed by us is as efficient as previous single photon
based scheme \cite{online-shop}. The entanglement based schemes are
more efficient than the single photon based schemes if densecoding
is exploited. However, the entanglement swapping based scheme (Protocol
3) proposed by us is least efficient as densecoding can not be exploited
in this case. On the other hand, this less efficient scheme has an
intrinsic advantage as message encoded qubits neither travel from
Alice to Bob nor are accessible to Eve. We have established the security
of all the proposed schemes from both individual attacks of an outsider
and participant attacks. We have not discussed the effect of noise
as the same on CDSQC has already been reported by us in case of non-Markovian
channels, which can be reduced to the Markovian channel in the limiting
case \cite{NM}. Further, a clear prescription for the study of the
effect noise is provided in (\cite{NM,crypt-switch,Vishal} and references
therein) using which single qubit (entangled state) based quantum
communication schemes are shown to perform better in the amplitude
and phase damping (collective noise) channels \cite{Vishal}. Following
the same strategy the effect of noise on the proposed protocols can
be studied with ease, and we expect Protocol 1 to perform better than
others over the amplitude and phase damping channels, while the other
schemes may be preferred when the qubits are transmitted in the collective
noise. 

We hope that the proposed alternatives of the quantum online shopping
schemes will provide experimentalists numerous possibilities to realize
the task in the near future.

\textbf{Acknowledgment:} KT and AP thank Defense Research \& Development
Organization (DRDO), India for the support provided through the project
number ERIP/ER/1403163/M/01/1603. They also thank Chitra Shukla for her interest in this work and some useful criticism of the work.


\begin{thebibliography}{10}
\bibitem{Shor-algo}Shor, P. W.: Polynomial-time algorithms for prime
factorization and discrete logarithms on a quantum computer. In Proc.
35th Annual Symp. on Foundations of Computer Science, (1994) Santa
Fe, IEEE Computer Society Press (1994)

\bibitem{bb84}Bennett, C. H., Brassard, G.: Quantum cryptography:
public key distribution and coin tossing. In: Proceedings of the IEEE
International Conference on Computers, Systems, and Signal Processing,
Bangalore, India, pp. 175-179 (1984)

\bibitem{ekert}Ekert, A.K.: Quantum cryptography based on Bell\textquoteright s
theorem. Phys. Rev. Lett. \textbf{67}, 661\textendash 663 (1991)

\bibitem{b92}Bennett, C.H.: Quantum cryptography using any two nonorthogonal
states. Phys. Rev. Lett. \textbf{68}, 3121-3124 (1992)

\bibitem{vaidman-goldenberg}Goldenberg, L., Vaidman, L.: Quantum
cryptography based on orthogonal states. Phys. Rev. Lett. \textbf{75},
1239-1243 (1995)

\bibitem{Long and Liu}Long, G. L., Liu, X. S.: Theoretically efficient
high-capacity quantum-key-distribution scheme. Phys. Rev. A \textbf{65},
032302 (2002)

\bibitem{ping-pong}Bostrom, K., Felbinger, T.: Deterministic secure
direct communication using entanglement. Phys. Rev. Lett. \textbf{89},
187902 (2002)

\bibitem{lm05}Lucamarini, M., Mancini, S.: Secure deterministic communication
without entanglement. Phys. Rev. Lett. \textbf{94}, 140501 (2005)

\bibitem{dsqcwithteleporta}Yan, F. L., Zhang, X. Q.: A scheme for
secure direct communication using EPR pairs and teleportation. Euro.
Phys. J. B\textbf{ 41}, 75-78 (2004)

\bibitem{reordering1}Zhu, A.D., Xia, Y., Fan, Q.B., Zhang, S.: Secure
direct communication based on secret transmitting order of particles.
Phys. Rev. A \textbf{73}, 022338 (2006)

\bibitem{With Anindita-pla}Banerjee, A., Pathak, A.: Maximally efficient
protocols for direct secure quantum communication. Phys. Lett. A \textbf{376},
2944-2950 (2012)

\bibitem{DLL}Deng, F.-G., Long, G. L., Liu, X.-S.: Two-step quantum
direct communication protocol using the Einstein-Podolsky-Rosen pair
block. Phys. Rev. A \textbf{68}, 042317 (2003)

\bibitem{With preeti}Yadav, P., Srikanth, R., Pathak, A.: Two-step
orthogonal-state-based protocol of quantum secure direct communication
with the help of order-rearrangement technique, \textbf{13}, 2731-2743
(2014) 

\bibitem{dsqc-ent swap}Shukla, C., Pathak, A.: Orthogonal-state-based
deterministic secure quantum communication without actual transmission
of the message qubits. Quantum Inf. Process. \textbf{13}, 2099-2113
(2014)

\bibitem{qd}Shukla, C., Kothari, V., Banerjee, A., Pathak, A.: On
the group-theoretic structure of a class of quantum dialogue protocols.
Phys. Lett. A \textbf{377}, 518 (2013)

\bibitem{switch}Srinatha, N., Omkar, S., Srikanth, R., Banerjee,
S., Pathak, A.: The Quantum Cryptographic Switch, Quantum Inf. Process.
\textbf{13}, 59-70 (2014)

\bibitem{cdsqc}Pathak, A.: Efficient protocols for unidirectional
and bidirectional controlled deterministic secure quantum communication:
Different alternative approaches. Quantum Inf. Process. \textbf{14},
2195 (2015)

\bibitem{my book}Pathak, A.: Elements of Quantum Computation and
Quantum Communication\emph{.} CRC Press, Boca Raton, USA (2013)

\bibitem{e-payment1}Wen, X.J.: An E-payment system based on quantum
group signature. Phys. Scr. \textbf{82}, 065403-065407 (2010) 

\bibitem{e-payment}Wen, X.J., Nie, Z.: An E-payment system based
on quantum blind and group signature. In: Proceedings of International
Symposium on Data, Privacy, and E-Commerce, America (2010)

\bibitem{interbank}Wen, X.J., Chen, Y.Z., Fang, J.B.: An inter-bank
E-payment protocol based on quantum proxy blind signature. Quantum
Inf. Process. \textbf{12}, 549-558 (2013)

\bibitem{interbank-cr-an}Cai, X.Q., Wei, C.Y.: Cryptanalysis of an
inter-bank E-payment protocol based on quantum proxy blind signature.
Quantum Inf. Process. \textbf{12}, 1651-1657 (2013)

\bibitem{chou}Chou, Y.-H., Lin, F.-J., Zeng, G.-J.: An efficient
novel online shopping mechanism based on quantum communication. Electronic
Commerce Research \textbf{14}, 349-367 (2014) 

\bibitem{online-shop}Huang, W., Yang, Y.-H., Jia, H.-Y.: Cryptanalysis
and improvement of a quantum communication-based online shopping mechanism.
Quantum Inf. Process. \textbf{14}, 2211-2225 (2015)

\bibitem{semi}Shukla, C., Thapliyal, K., Pathak, A.: Semi-quantum
communication: protocols for key agreement, controlled secure direct
communication and dialogue, Quantum Inf. Process. \textbf{16}, 295
(2017)

\bibitem{proxy-1}Shao, A.-X., Zhang, J.-Z., Xie, S.-C.: An e-payment
protocol based on quantum multi-proxy blind signature. Int. J. Theor.
Phys. \textbf{56}, 1241-1248 (2017) 

\bibitem{blind}Zhang, J.-Z., Yang, Y.-Y., Xie, S.-C.: A third-party
e-payment protocol based on quantum group blind signature. Int. J.
Theor. Phys. \textbf{56}, 2981-2989 (2017) 

\bibitem{New-ref-1}Buscemi, F., Hall, M. J., Ozawa, M., Wilde, M.
M.: Noise and disturbance in quantum measurements: an information-theoretic
approach. Phys. Rev. Lett. \textbf{112}, 050401 (2014)

\bibitem{New ref 2}Biham, E., Boyer, M., Boykin, P. O., Mor, T.,
Roychowdhury, V.: A proof of the security of quantum key distribution.
Journal of Cryptology \textbf{19}, 381-439 (2006)

\bibitem{With chitra IJQI}Shukla, C., Pathak, A., Srikanth, R.: Beyond
the Goldenberg-Vaidman protocol: secure and efficient quantum communication
using arbitrary, orthogonal, multi-particle quantum states. Int. J.
Quantum Inf. \textbf{10}, 1241009 (2012)

\bibitem{nielsen}Nielsen, M.A., Chuang I. L.: Quantum Computation
and Quantum Informatiom. Cambridge University Press, New Delhi, 589
(2008)

\bibitem{decoy}Sharma, R. D., Thapliyal, K., Pathak, A., Pan, A.
K., De, A.: Which verification qubits perform best for secure communication
in noisy channel? Quantum Inf. Process. \textbf{15}, 1703-1718 (2016)

\bibitem{crypt-switch}Thapliyal, K., Pathak, A.: Applications of
quantum cryptographic switch: Various tasks related to controlled
quantum communication can be performed using Bell states and permutation
of particles. Quantum Inf. Process. \textbf{14}, 2599 (2015) 

\bibitem{Vishal}Sharma, V., Thapliyal, K., Pathak, A., Banerjee,
S.: A comparative study of protocols for secure quantum communication
under noisy environment: single-qubit-based protocols versus entangled-state-based
protocols. Quantum. Inf. Process. \textbf{15}, 4681 (2016)

\bibitem{NM}Thapliyal, K., Pathak, A., Banerjee, S.: Quantum cryptography
over non-Markovian channels. Quantum Inf. Process. \textbf{16}, 115
(2017) 

\bibitem{AQD}Banerjee, A., Shukla, C., Thapliyal, K., Pathak, A.,
Panigrahi, P. K.: Asymmetric quantum dialogue in noisy environment.
Quantum Inf. Process. \textbf{16}, 49 (2017)

\bibitem{QC}Banerjee, A., Thapliyal, K., Shukla, C., Pathak, A.:
Quantum Conference, Quantum Inf. Process. \textbf{17}, 161 (2018)

\bibitem{QInternet}Kimble, H. J.: The quantum internet. Nature \textbf{453},
1023-1030 (2008)

\bibitem{QInternet2}Pirandola, S., Braunstein, S. L.: Physics: Unite
to build a quantum Internet. Nature \textbf{532}, 169-171 (2016)

\bibitem{Voting1}Thapliyal, K., Sharma, R. D., Pathak, A.: Protocols
for quantum binary voting. Int. J. Quantum Info. \textbf{15}, 1750007
(2017)

\bibitem{Our-auction}Sharma, R. D., Thapliyal, K., Pathak, A.: Quantum
sealed-bid auction using a modified scheme for multiparty circular
quantum key agreement. Quantum Inf. Process. \textbf{16}, 169 (2017)

\bibitem{Cor-eli}Song, J., Zhang, S.: Comment on: \textquotedbl{}Quantum
exam\textquotedbl{}. Phys. Lett. A \textbf{350}, 174 (2006)

\bibitem{authentication1}Kanamori, Y., Yoo, S.M., Gregory, D.A.,
Sheldon, F.T.: On quantum authentication protocols. InGLOBECOM'05.
IEEE Global Telecommunications Conference, IEEE. \textbf{3}, 5 (2005)

\bibitem{authentication2}Ljunggren, D., Bourennane, M., Karlsson,
A.: Authority-based user authentication in quantum key distribution.
Phys. Rev. A. \textbf{62}, 022305 (2000)

\bibitem{Rev-cryp}Gisin, N., Ribordy, G., Tittel, W., Zbinden, H.:
Quantum cryptography. Rev. Mod. Phys. \textbf{74}, 145 (2002) 

\bibitem{Troj-hor}Deng, F.-G., Li, X.-H., Zhou, H.-Y., Zhang, Z.-j.:
Improving the security of multiparty quantum secret sharing against
Trojan horse attack. Phys. Rev. A \textbf{72}, 044302 (2005)

\bibitem{Troj-hor1}Li, X.-H., Deng, F.-G., Zhou, H.-Y.: Improving
the security of secure direct communication based on the secret transmitting
order of particles. Phys. Rev. A \textbf{74}, 054302 (2006) 

\bibitem{defn of qubit efficiency}Cabello, A.: Quantum key distribution
in the Holevo limit. Phys. Rev. Lett. \textbf{85}, 5635\textendash 5638
(2000)

\bibitem{QPC}Thapliyal, K., Sharma, R. D., Pathak, A.: Orthogonal-state-based
and semi-quantum protocols for quantum private comparison in noisy
environment. arxiv:1608.00101v1 (2016) 
\end{thebibliography}
\end{document}